%% file: HNL_PRL_Accepted_v1.tex
\documentclass[%
reprint,
superscriptaddress,
longbibliography,
amsmath,
amssymb,
aps,
prl,
floatfix]{revtex4-2}

\usepackage{xcolor}
\usepackage{graphicx}
\usepackage{dcolumn}
\usepackage{bm}
\usepackage[pdftex]{hyperref} 
\usepackage[mathlines]{lineno}
\usepackage{enumerate}%

\usepackage{svg}

\usepackage{adjustbox}
\usepackage{multirow}
\usepackage{lineno}

\newcommand{\mpi}{\mu\pi}
\newcommand{\ee}{\nu e^+e^-}
\newcommand{\vpi}{\nu\pi^0}

\newcommand{\mumix}{\lvert U_{\mu 4}\rvert^2}

\def\mhnl{{\ensuremath{m_{\mathrm{HNL}}}}}

\begin{document}

\preprint{Fermilab-XXX}

\widetext


\title{Search for heavy neutral leptons in electron-positron and neutral-pion final states with the MicroBooNE detector}

\input{microboone-author-list-september2023-PRD}

\date{30 November 2023}

\begin{abstract}
We present the first search for heavy neutral leptons (HNL) decaying into $\ee$ or $\vpi$ final states in a liquid-argon time projection chamber
using data collected with the MicroBooNE detector. The data were recorded synchronously with the NuMI neutrino beam from Fermilab's Main Injector corresponding to a total exposure of $7.01 \times 10^{20}$ protons on target.
We set upper limits at the $90\%$ confidence level on the mixing parameter $\lvert U_{\mu 4}\rvert^2$ in the mass ranges $10\le m_{\rm HNL}\le 150$~MeV for the $\ee$ channel and
$150\le m_{\rm HNL}\le 245$~MeV for the $\vpi$ channel,
assuming $\lvert U_{e 4}\rvert^2 = \lvert U_{\tau 4}\rvert^2 = 0$.
These limits represent the most stringent constraints in the mass range $35<\mhnl<175$~MeV and the first constraints from a direct search for $\vpi$ decays.
\end{abstract}

\maketitle

Heavy neutral leptons (HNL) appear in minimal extensions of the standard model (SM) that can explain the origin of neutrino masses, the generation of the baryon asymmetry through leptogenesis, and the nature of dark matter~\cite{Abdullahi:2022jlv}. 
They are introduced through an extension of the Pontecorvo-Maki-Nakagawa-Sakata (PMNS) matrix by adding heavy mass eigenstates that mix very weakly with the three active neutrino states. For a single HNL state, the extended PMNS matrix has the dimension $4\times 4$, which leads to four new parameters: the HNL mass $\mhnl$ and three mixing parameters, $\lvert U_{\alpha 4}\rvert^2$ with $\alpha$ = $e$, $\mu$, or $\tau$. 
The HNL production and decay rates are suppressed by the elements $|U_{\alpha4}|^2$ through mixing-mediated interactions with SM gauge bosons. 
A vibrant experimental program is dedicated to searching for HNLs and other feebly-interacting particles~\cite{Antel:2023hkf}.

Here, we use data recorded with the MicroBooNE detector to perform a search for HNLs decaying to $\ee$ or $\vpi$ final states. The MicroBooNE detector~\cite{Acciarri:2016smi} is one of the three liquid-argon time projection chambers (LArTPC) comprising the Fermilab short-baseline neutrino program~\cite{Machado:2019oxb}. The liquid-argon technology
provides a powerful tool to search for signatures of physics beyond the SM as it allows us to fully reconstruct decays 
through its precision imaging capability. 
Since the two final states, $\ee$ and $\vpi (\pi^0\to\gamma\gamma$),
are topologically very similar, leading to two electromagnetic showers in the LArTPC, 
the search is performed within a single analysis framework using boosted decision trees (BDTs).
This analysis strategy is based on our previous searches for HNL decays to $\mu\pi$ final states~\cite{MicroBooNE:2022ctm} and decays of Higgs portal scalars into $e^+e^-$ pairs~\cite{MicroBooNE:2021usw}.

The MicroBooNE detector recorded data between 2015 and 2021.
It was simultaneously exposed on-axis to the booster neutrino beam (BNB)~\cite{AguilarArevalo:2008yp} and
off-axis to the neutrino beam from the main injector (NuMI)~\cite{ADAMSON2016279}. 
Only NuMI data are used for this search, since the higher average beam energy compared to the BNB leads to a higher kaon rate and therefore 
potentially more HNL production.

\begin{figure}[ht!]
\centering
\includegraphics[width=0.45\textwidth]{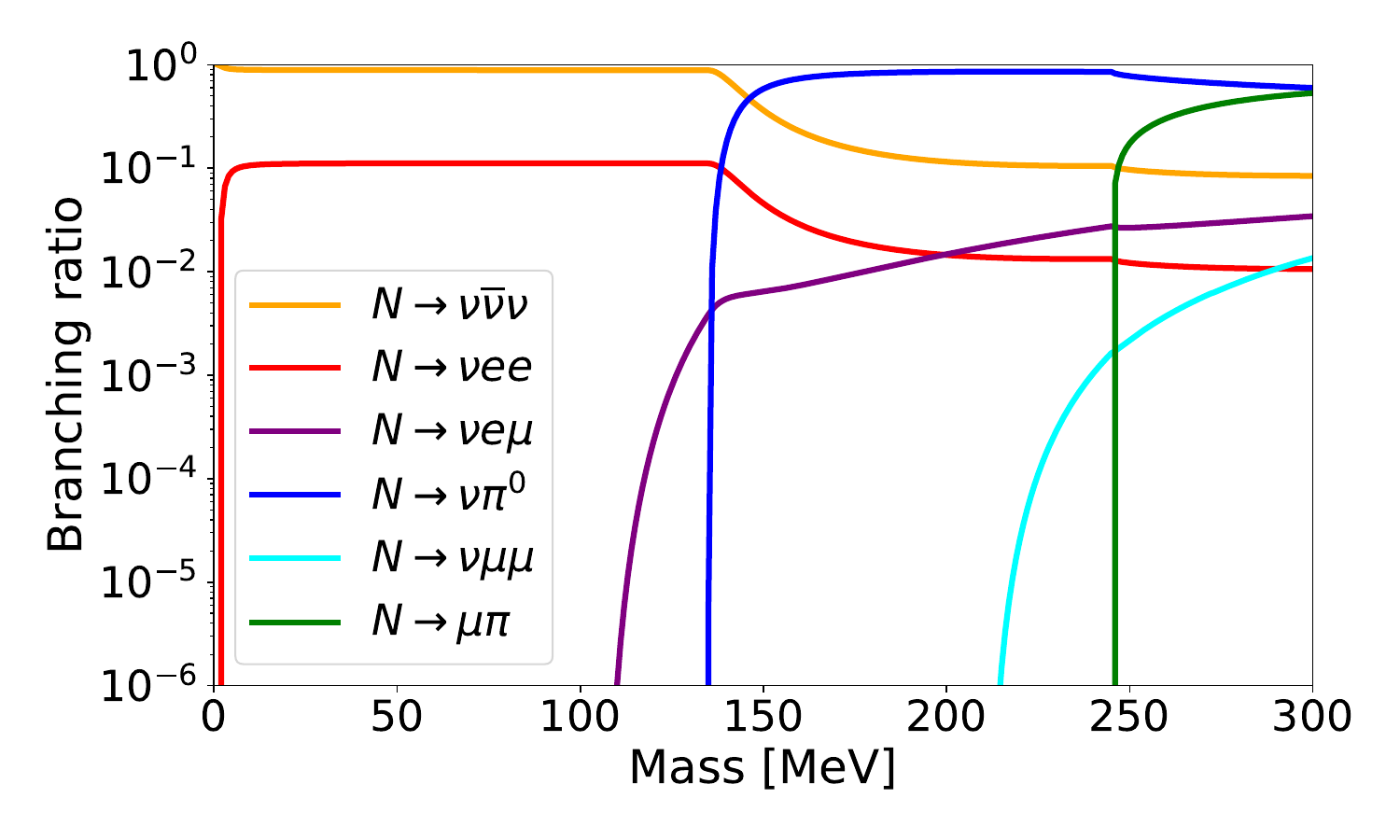}
\caption{Branching ratios for Majorana HNL decays with $\mumix>0$ in the range $0\le\mhnl\le300$~MeV calculated with the equations of Ref.~\protect\cite{Ballett:2019bgd},
assuming $\lvert U_{e 4}\rvert^2 = \lvert U_{\tau 4}\rvert^2 = 0$. Both conjugations of charged leptons are included in the relevant channels. } 
\label{fig:Branching_ratios}
\end{figure}

We assume that the HNLs are produced in the absorber, made from aluminium, steel, and concrete, which is
located $\approx 725$~m downstream from the NuMI beam's graphite target and $\approx 104$~m from the MicroBooNE detector~~\cite{ADAMSON2016279}. 
The absorber is located downstream of the
MicroBooNE detector at the end of the NuMI decay pipe.
HNLs produced in the absorber would approach the detector in
almost the opposite direction to the neutrinos that originate from the NuMI beam target, 
which significantly improves background rejection~\cite{MicroBooNE:2022ctm}.
Approximately $13\%$ of the beam protons reach the absorber and produce
$K^+$ mesons that can decay at rest into HNLs through the process $K^+\to \mu^+N$, 
while most of the $K^-$ mesons are absorbed. 
If the HNL lifetime is sufficiently long, the HNLs could reach the MicroBooNE detector and decay into SM particles within the argon.
The sensitivity of MicroBooNE to this production mechanism has previously been studied in Ref.~\cite{Kelly:2021xbv}.

The kinematic distributions of the final state particles in the HNL decay depend on \mhnl, the kinetic energy of the HNL, and whether the HNL is assumed to be a Dirac or Majorana particle. We encode the production and decay properties of the HNL using the equations in Ref.~\cite{Ballett:2019bgd} in a simulation code developed for
MicroBooNE's previous HNL search~\cite{MicroBooNE:2022ctm}. We also validated the simulation with a recent implementation of HNL kinematics in the {\sc GENIE} generator~\cite{Plows:2022gxc}.
The branching ratios
for $\mhnl<300$~MeV are shown in Fig.~\ref{fig:Branching_ratios}. 
We assume $\lvert U_{e 4}\rvert^2 = \lvert U_{\tau 4}\rvert^2 = 0$ as $\lvert U_{e 4}\rvert^2$ is already
severely constrained~\cite{Antel:2023hkf} and $\lvert U_{\tau 4}\rvert^2$ is 
not kinematically accessible. 
Neglecting the ``invisible" decay $N\to 3\nu$, the decays into 
$\ee$ final states dominate for $\mhnl$ values below the mass
of the $\pi^0$ meson, and the $\vpi$ final states 
dominate above. We generate samples for
different $\mhnl$, and in the $\ee$ and $\vpi$ final states, to cover the full
range of accessible model parameters.

We use NuMI data corresponding to $7.01 \times 10^{20}$~protons on target (POT), which were taken in two operating modes, forward horn current 
(FHC) with $2.00 \times 10^{20}$~POT (Run~1) and reverse horn current (RHC) with $5.01 \times 10^{20}$~POT (Run~3). 
The two data sets are analyzed separately to account for differences in neutrino flux and detector configuration. We assume equal rates of $K^+$ production for the two horn polarities~\cite{MiniBooNEKDAR}. 

We select a ``beam-on" data sample to search for an HNL signal where the event triggers coincide with the NuMI beam. 
Such beam-on events are frequently triggered by a cosmic ray and not a neutrino interaction. This type of event is modeled by selecting a ``beam-off" sample collected under identical trigger conditions but when no neutrino beam is present. 
 The ``beam-off" sample is normalized to the number of triggers recorded in the beam-on
data.
Neutrino-induced background from the NuMI beam is modeled using a 
Monte Carlo (MC) simulation~\cite{Andreopoulos:2015wxa}, with
cosmic rays and noise from data overlaid on the simulation.
The ``in-cryostat $\nu$" sample contains interactions of neutrinos with the argon inside the cryostat, and 
the ``out-of-cryostat $\nu$" sample describes interactions with the detector structure and surrounding material.  Both samples are normalized to the numbers of POT of the data sample. An additional data-driven scaling factor is applied to the out-of-cryostat $\nu$ sample.

We reconstruct neutrino interactions and cosmic rays within the argon with a chain of pattern-recognition algorithms, implemented using the \texttt{Pandora} Software Development Kit (SDK)~\cite{Marshall:2015rfa,MicroBooNE:2017xvs}.
Hits are formed from the waveforms
read out by three anode wire planes -- two induction planes and one charge collection plane. We then group hits into slices
to isolate neutrino interactions and cosmic rays. 
Slices are reconstructed under
both hypotheses, and 
a support vector machine then calculates a ``topological score" to classify slices as either a neutrino interaction or cosmic ray.
We select events with exactly one neutrino slice to examine them
for candidate HNLs.

``Objects" are then reconstructed either as a track, as expected for a minimum ionizing particle, or a shower, consistent with being an electron or photon. The distinction between tracks and showers is performed
using a ``track score" that mainly relies on the profile of the charge deposition, the range, and topological information. 

The start and end points of all objects associated with the slice must lie within the TPC's fiducial 
volume~\footnote{The MicroBooNE coordinate system is right-handed. The $x$ axis points along the negative drift direction, the $y$
 axis vertically upward, and the $z$ axis
 along the direction of the BNB beam. The polar angle $\theta$ is defined
 with respect to the $z$ axis, and the azimuthal angle $\phi$ with respect to the $x$ axis. The $yz$ plane is the charge collection plane. 
 The TPC's fiducial volume is defined by $9<x<253$~cm,
 $|y|<112$~cm, and $14<z<1020$~cm}, 
 and the fraction of reconstructed hits in the slice contained within the fiducial volume has to be $>0.9$.
 The energy of all the objects in the slice, $E_{\textrm{sl}}$, is reconstructed from the charge read-out on the TPC's charge collection plane. 
 We require $E_{\textrm{sl}}<500$~MeV as the energy deposited from the decays of HNLs with a mass $\mhnl<245$~MeV is expected to be lower than for most beam or cosmic-ray events. 
 We require $E_{\textrm{sl}}<500$~MeV as the decays of HNLs with a mass $\mhnl<245$~MeV are expected to deposit less energy than most neutrino or cosmic-ray interactions. 
 
Light flashes are reconstructed from the waveforms of an array 
of 32 photomultiplier tubes.  
We require that the time of the 
largest flash in a 23~$\mu$s window surrounding the NuMI beam trigger coincides with the NuMI beam spill of $\approx 10 \mu$s.
A cosmic ray tagger (CRT) surrounding the cryostat was installed about midway through MicroBooNE 
operations~\cite{MicroBooNE:2019lta}.
If there is a hit recorded by the CRT within $1$~$\mu$s of the flash for the Run~3 (RHC) sample, the event is identified as a cosmic ray and is rejected~\cite{MicroBooNE:2020akw}.
The CRT is not used for the Run~1 (FHC) data set as it was not yet operational at that time.
To further reduce cosmic-ray background, we require the ``flash match score" to be $<15$. 
This is calculated as a $\chi^2$ value 
by comparing the light signals in the PMTs  
to the expected  PMT signals assuming the recorded charge is due to a neutrino interaction.

\begin{table}[htb]
\centering
\caption{Numbers of events that remain after preselection normalized to the POT for the two data samples.
The percentages are the contributions of each sample to the sum of the background predictions.
}
\setlength{\tabcolsep}{5pt} 
\begin{tabular}{lcc}
\hline\hline 
Sample & Run 1 (FHC) & Run 3 (RHC) \\
POT  & $2.00 \times 10^{20}$  & $5.01 \times 10^{20}$  \\
 \hline 
Beam-off & $3548$ $(46\%) $  &  $3597$ $(33\%)$\\
In-cryostat $\nu$ & $3607$ $(47\%) $  &   $6805$ $(63\%)$  \\
Out-of-cryostat $\nu$ & 
$\phantom{0}567$ \phantom{0}$(7\%)$  
& $\phantom{0}464$ \phantom{0}$(4\%)$\\
\\
Sum of predictions & $7722$ & $10866$ \\
Beam-on (data)  & $7598$ &  $11282$ \\\\
Data over prediction & $0.98$ & $1.04$ \\
\hline\hline
\end{tabular}
\label{tab:presel}
\end{table}

\begin{figure*}[htb!]
   \centering
\mbox{ \includegraphics[width=0.33\textwidth]{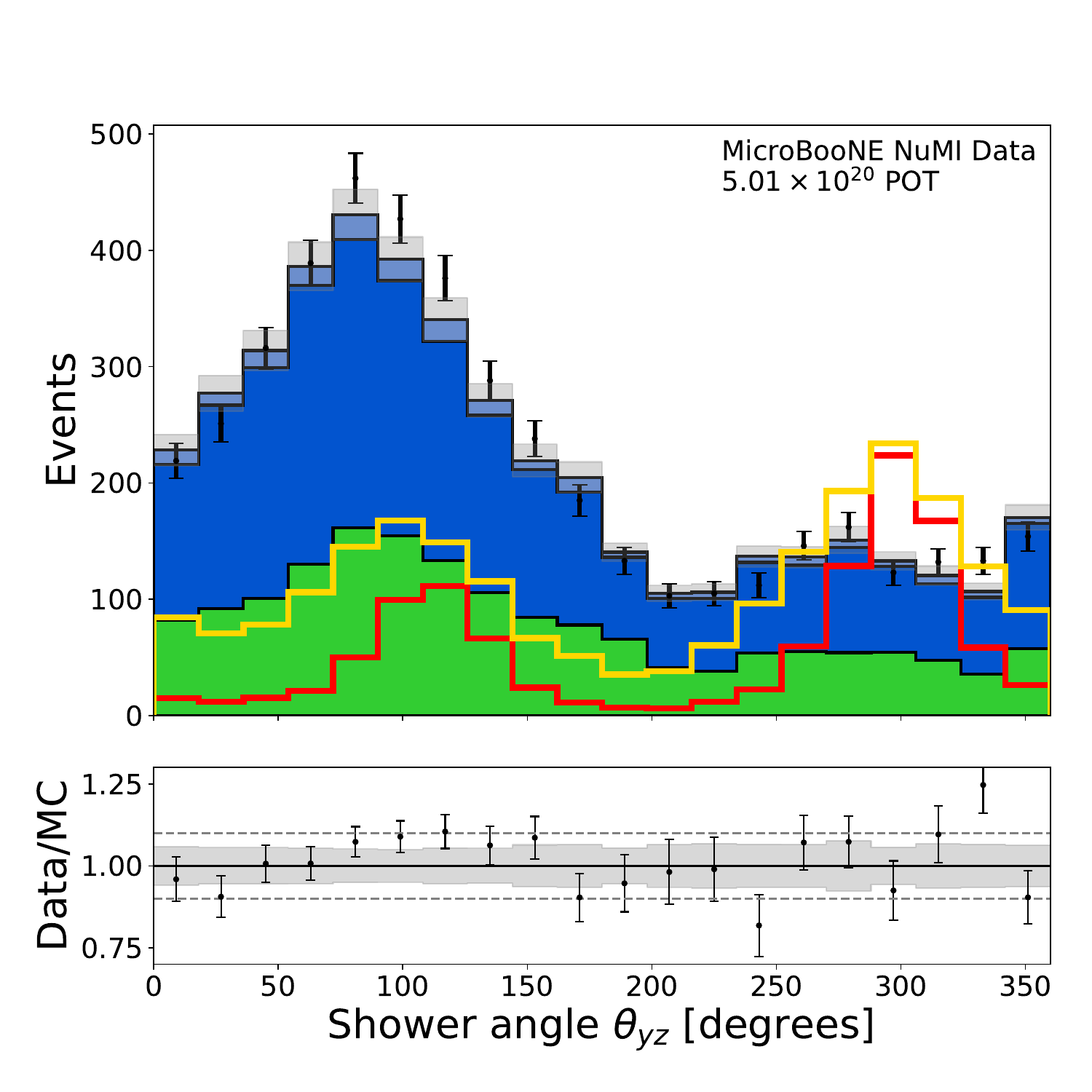}
 \includegraphics[width=0.33\textwidth]{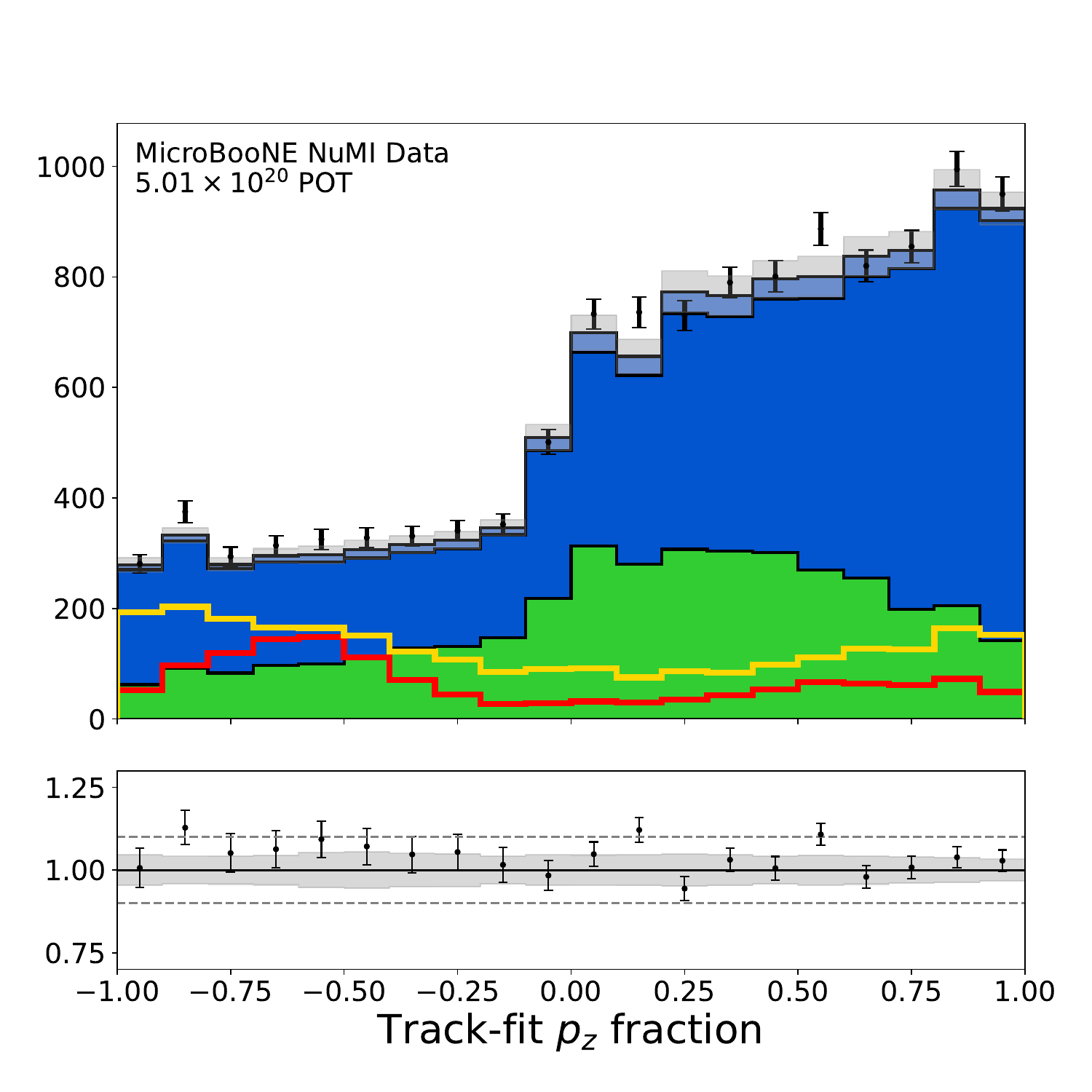}
\includegraphics[width=0.33\textwidth]{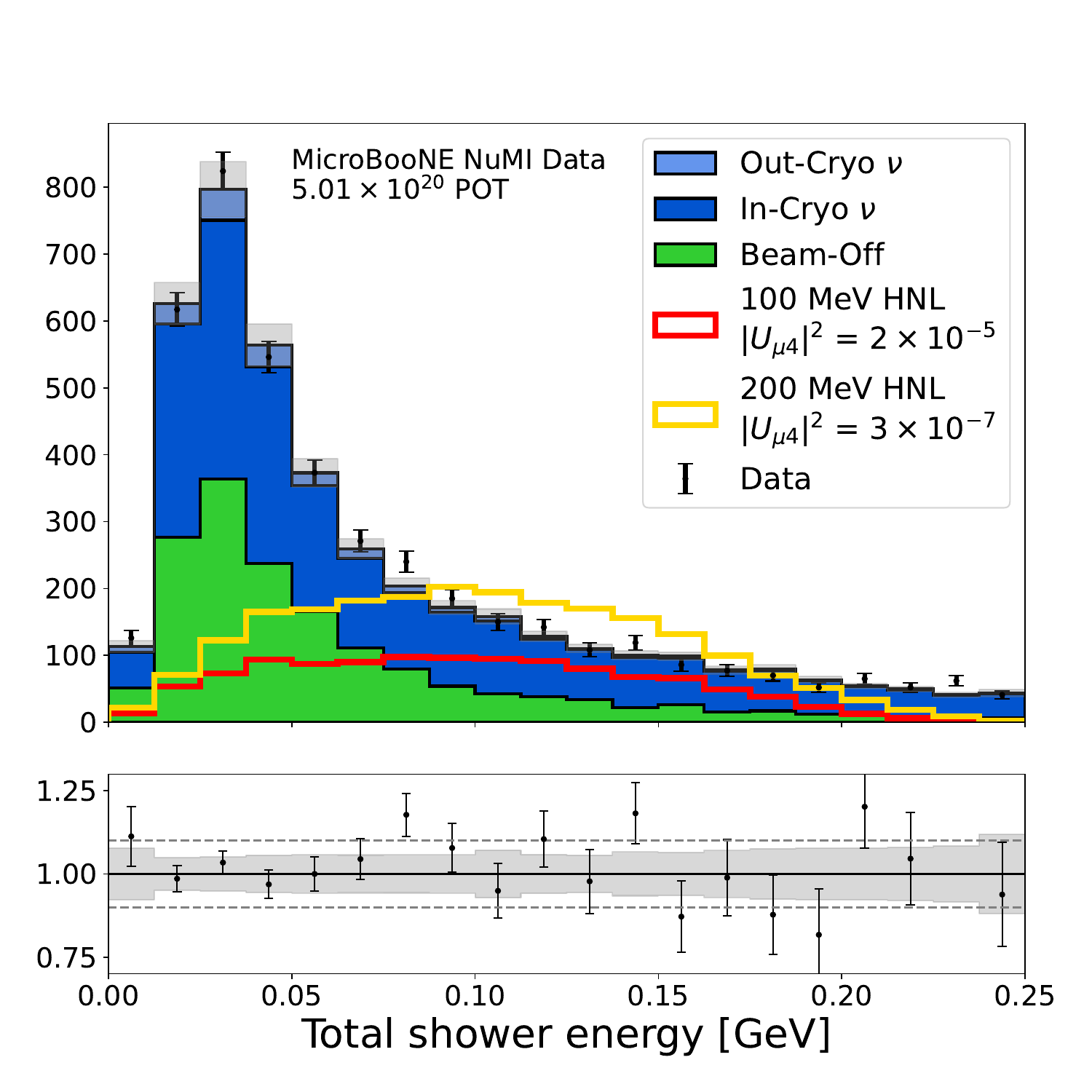}}
\caption{BDT input variables after the preselection for Run 3 (RHC) data: (a) shower angle $\theta_{yz}$ with respect to the $y$ axis projected on the $yz$ plane, (b) track-fit $z$ momentum fraction, and (c) the total shower energy for data and the background prediction. The signal distributions for $N\to\ee$ decays with $\mhnl=100$~MeV and $N\to\vpi$ decays with $\mhnl=200$~MeV are
normalized to $\mumix=2\times 10^{-5}$ and $\mumix=3\times 10^{-7}$, respectively.
The gray band indicates the quadrature sum of all uncertainties on the background expectation.}  
   \vspace{-69mm}
 \makebox[\textwidth][l]{\hspace{1.1cm}(a) \hspace{5.2cm} (b) \hspace{6.2cm} (c)}\\ \vspace{69mm}
\label{fig:BDT_input}
\end{figure*}

Table~\ref{tab:presel} shows the effect of the preselection requirements for the background samples. 
The signal efficiency after preselection is $\approx 35\%$ for the different $\mhnl$ and final states, while
we retain $\approx4\%$ of the in-cryostat neutrino interactions. 
The contribution of the beam-off and out-of-cryostat $\nu$ events to the background sample is significantly smaller for Run 3 compared to Run 1, since the CRT improves the rejection of these classes of events.
 The numbers of data events after the preselection agree well, within $(2$--$4)\%$, with the sum of the predictions from the three main background sources.

At this stage of the selection, we use \texttt{XGBoost}~\cite{xgboost} to train BDTs 
that optimize the
discrimination between signal and background in this selected sample. 
A separate training is performed for each mass point, final state, and data
set (FHC/RHC) using subsets of the signal sample and the three background samples. The training 
samples are excluded from the subsequent analysis.
We reduce the number of input variables to $20$ from a potential set of several hundred variables by training BDT models on the full set of variables and then identifying the variables with
the largest impact that are common to all tested HNL model parameters. 

As variables defined for the event (slice), we use
the multiplicity of objects, 
the track multiplicity,
the total energy measured from all tracks, 
the total energy measured from all showers, the total energy, 
 the energy of the highest energy track,
 the topological and flash match scores,
 the energy deposited in the first $4$~cm of the highest energy shower,
 and the shower angle $\theta_{yz}$, which is the average direction of all showers calculated with respect to the $y$ axis projected onto the $yz$ plane.

We also use the number of hits on each of the three wire planes associated with the highest-energy object and the object's track score, where the
highest energy object is determined 
by the associated number of hits on the wires of the collection plane.

The final set of variables use angular information for the highest energy object: 
the polar angles, the azimuthal angles, and the $z$-momentum fractions, calculated as
 the ratio of the momentum component in the $z$ coordinate over the total object momentum.
 These angular variables are calculated twice for each object by fitting it as track and as a shower.

If there is no track or shower, some variables are treated as missing in the BDT by using a placeholder. The distributions in Fig.~\ref{fig:BDT_input} show the shower angle $\theta_{yz}$, the track-fit $z$ momentum fraction, and the total shower energy for data and the background prediction in Run~3. 
These variables were found, during BDT training procedures, to be amongst the most sensitive to an HNL signal.
Momenta of particles produced in neutrino interactions predominantly point in the $+z$ direction, whereas signal is more clustered around $-z$.
The shower angle for signal has two peaks, depending on whether the start and end point of the shower are correctly identified. 
Since the BDT uses the information of all variables, such incorrectly reconstructed
events can still be identified as signal.

\begin{figure}[htbp]
\centering
\includegraphics[width=0.4\textwidth]{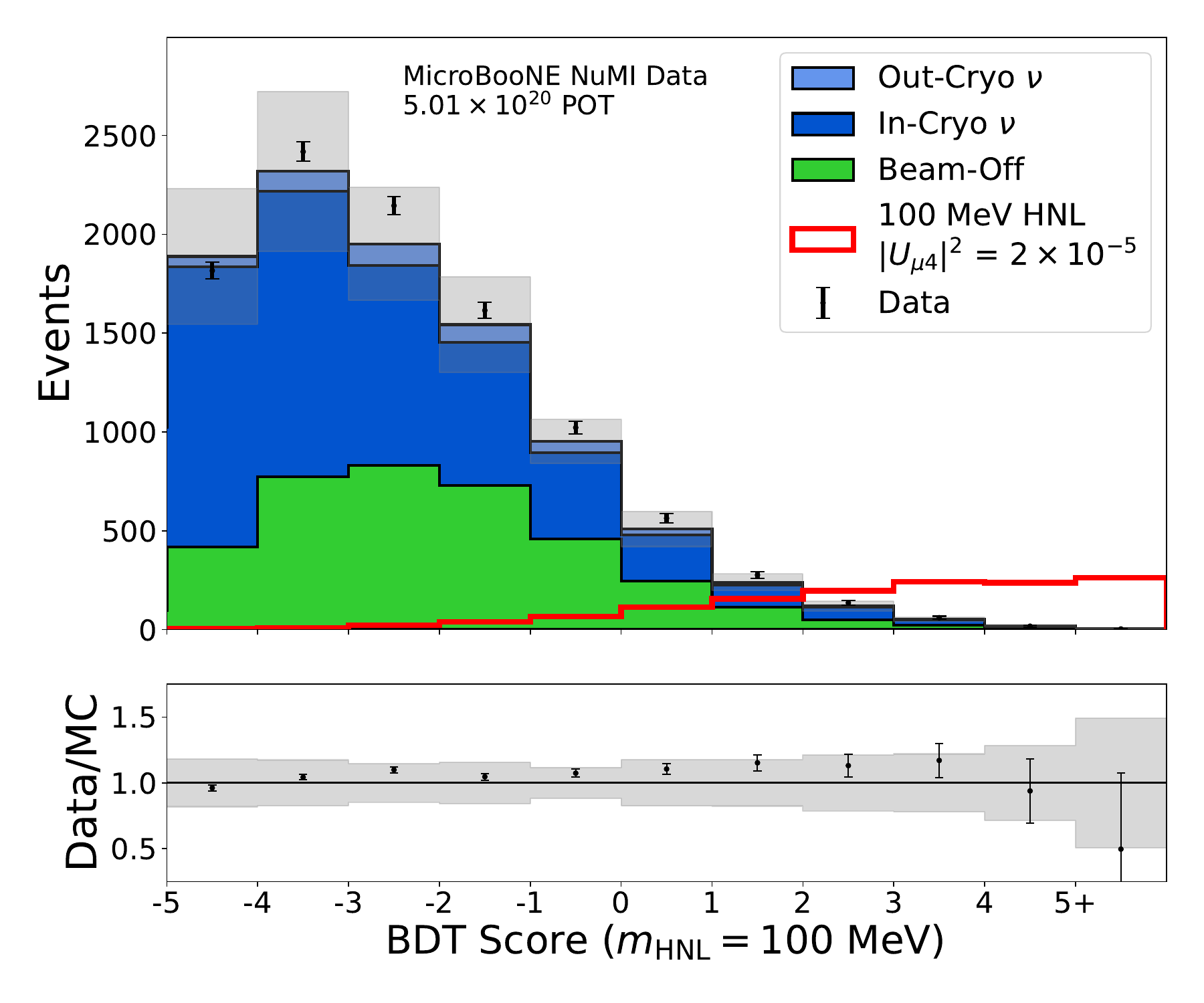}
\caption{BDT score distribution for the model trained with $N\to\ee$ decays at $\mhnl=100$~MeV, compared to Run~3 (RHC) data. The
signal distribution is normalized to $\mumix=2\times 10^{-5}$. The gray band indicates the quadrature sum of all uncertainties on the background expectation.} 
\label{fig:BDT_output_data_run3_ee}
\end{figure}

The background contributing to the 
BDT score distribution shown in
Fig.~\ref{fig:BDT_output_data_run3_ee}
is expected to be dominated by in-cryostat $\nu$ interactions. 
The BDT identifies and rejects most charged-current $\nu_{\mu}$ interactions.
For BDT scores $>3$, about $40\%$ of the simulated in-cryostat $\nu$ events are neutral-current interactions producing $\pi^{0}$ mesons, as this topology resembles the $N\to\ee$ and $N\to\vpi$ decays.

To determine the sensitivity to a possible HNL signal, 
we evaluate systematic uncertainties that could modify the BDT score distributions for signal and background~\cite{DavidMarsden}.
For the in-cryostat $\nu$ background, we consider the impact of the flux simulation, the neutrino-argon cross-section modeling, hadron interactions with argon, and detector modeling.
The beam-off sample is taken from data and therefore has no associated systematic uncertainties other than the statistical fluctuations in the sample. The impact of the normalization uncertainty on the out-of-cryostat 
$\nu$ sample is negligible, as the contribution to the final sample is small~\cite{MicroBooNE:2022ctm}.

The dominant uncertainty on the background in the signal region at high BDT scores is due to the statistical uncertainty of the samples, since most of the background has been rejected. We therefore extrapolate
systematic uncertainties from higher-statistics regions of the BDT score
distribution to the signal region. The quadrature sum of the background detector modelling uncertainty is taken to be $30\%$. 

The dominant contribution to the systematic uncertainty on the signal sample arises from the rate of kaon production at rest in the NuMI absorber. It is taken to be $\pm 30\%$ based on the evaluation by the MiniBooNE collaboration~\cite{MiniBooNEKDAR}. The sum of the detector-related systematic uncertainties is $(10-20)\%$.
The systematic uncertainties are separately evaluated for all signal parameters used in the BDT training, with consistent results. Due to the higher number of POTs and the better cosmic-ray rejection of the CRT, the signal sensitivity is dominated by the Run~3 data set.

The BDT score distributions are used to derive limits on
$\mumix$ for the different model parameters. We use the \texttt{pyhf} algorithm~\cite{pyhf_paper}, which
is an implementation of a statistical model to estimate confidence intervals for multi-bin histograms, based on the asymptotic formulas of Ref.~\cite{Cowan:2010js}.
The formalism allows for the treatment of systematic uncertainties through the use of profile likelihood ratios.
The results are validated with the modified frequentist CL$_s$ calculation of the \texttt{COLLIE} program~\cite{Fisher:2006zz}.
The \texttt{pyhf} code scans over a range of scaling parameters for the signal normalization and returns an interpolated value of 
the scaling parameter that corresponds to the $90\%$ confidence level (CL$_s=0.1$). 
The BDT distributions for each run period (Run~1 and Run~3) enter the limit setting as separate channels before their likelihoods are combined. The statistical uncertainties on signal and background are uncorrelated, whereas the systematic uncertainties on the flux and 
out-of-cryostat $\nu$ normalization are taken as fully correlated between
the run periods.
We studied the impact of the other systematic uncertainties with different assumptions about their correlations and determined that this choice has only a small impact on the result.
Using BDT models
trained with neighboring mass points shows no significant deterioration of sensitivity.  

\begin{table}[htbp]
\caption{The $90\%$~CL observed and median expected limits on $\mumix$ as a function of $\mhnl$ assuming a Majorana state.}
\setlength{\tabcolsep}{6pt}
\begin{tabular}{cccc}
   \hline  \hline
$\mhnl$   & \multicolumn{3}{c}{Limit $\mumix$}\\
(MeV)   & Observed &  Median & Standard deviation
              \\
\hline
$\ee$ & \multicolumn{3}{c}{}\\
$10$ & $3.32\times10^{{-3}}$ & $3.59\times10^{{-3}}$ & $2.92$--$4.52\times10^{{-3}}$ \\ 
$20$ & $3.82\times10^{{-4}}$ & $3.94\times10^{{-4}}$ & $3.18$--$5.02\times10^{{-4}}$ \\ 
$50$ & $2.96\times10^{{-5}}$ & $2.86\times10^{{-5}}$ & $2.36$--$3.55\times10^{{-5}}$ \\ 
$100$ & $2.94\times10^{{-6}}$ & $3.16\times10^{{-6}}$ & $2.61$--$3.92\times10^{{-6}}$ \\ 
$150$ & $5.99\times10^{{-7}}$ & $7.71\times10^{{-7}}$ & $6.30$--$9.71\times10^{{-7}}$ \\
\hline
$\vpi$ & \multicolumn{3}{c}{}\\
$150$ & $2.15\times10^{{-7}}$ & $2.25\times10^{{-7}}$ & $1.86$--$2.79\times10^{{-7}}$ \\ 
$180$ & $7.01\times10^{{-8}}$ & $6.87\times10^{{-8}}$ & $5.64$--$8.56\times10^{{-8}}$ \\ 
$200$ & $3.95\times10^{{-8}}$ & $4.43\times10^{{-8}}$ & $3.65$--$5.51\times10^{{-8}}$ \\ 
$220$ & $3.97\times10^{{-8}}$ & $3.62\times10^{{-8}}$ & $3.00$--$4.46\times10^{{-8}}$ \\ 
$240$ & $2.67\times10^{{-8}}$ & $2.71\times10^{{-8}}$ & $2.23$--$3.36\times10^{{-8}}$ \\ 
$245$ & $2.26\times10^{{-8}}$ & $2.29\times10^{{-8}}$ & $1.87$--$2.85\times10^{{-8}}$ \\ 
\hline\hline
\end{tabular}
\label{tab:HNLResults}
\end{table}

The observed limits for the model parameters tested, given in Table~\ref{tab:HNLResults}, are all within $2$ standard deviations of the median expected limit. 
A linear interpolation is performed between the tested $\mhnl$ hypotheses, which slightly underestimates the sensitivity in the interpolation regions.

\begin{figure*}[htbp]
  \centering
  \includegraphics[width=0.80\textwidth]{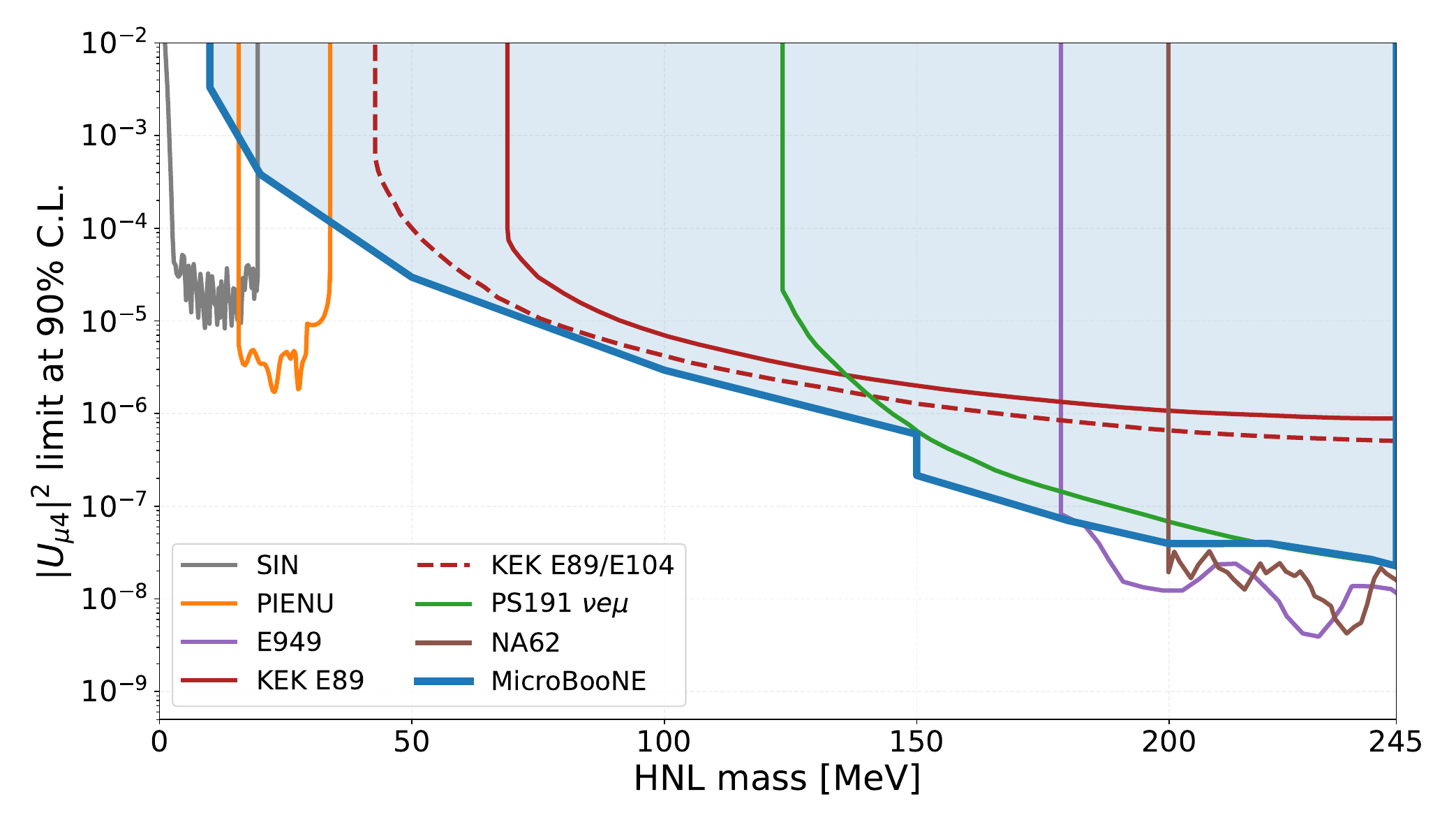}
  \caption{
  Limits on $\mumix$ at the $90\%$ CL assuming $\lvert U_{e 4}\rvert^2 = \lvert U_{\tau 4}\rvert^2 = 0$ as a function of mass for a Majorana HNL in the ranges $10<\mhnl<150$~MeV 
  ($N\to\ee$) and $150<\mhnl<245$~MeV ($N\to\vpi$).
    We use a linear interpolation between the tested $\mhnl$ hypotheses. 
    The discontinuity at 
  $\mhnl=150$~MeV is due to the change in decay channel from
  $\ee$ to $\vpi$.
  The constraints are compared to the
  published results of the SIN~\protect\cite{Daum:1987bg}, PIENU~\protect\cite{PIENU:2019usb}, KEK-E89~\protect\cite{Hayano:1982wu}, BNL-E949~\protect\cite{E949_limit_2015}, NA62~\protect\cite{NA62:2021bji},
  and PS191~\protect\cite{PS191_1987} collaborations. 
  The unpublished limit using the KEK-E89/E104 data~\protect\cite{Yamazaki:1984sj} is shown as a dashed line. 
  }
  \label{fig:limits}
\end{figure*}

We also consider Dirac HNL states. 
The Dirac HNL decay rate is a factor of 2 smaller than for Majorana states at the same value of $\mumix$.
 Since the effect of differing decay kinematics on the sensitivity is found to be negligible, the limits for Dirac states can be obtained by scaling the results in
 Table~\ref{tab:HNLResults} by a factor of $\sqrt{2}$.

We compare our results for Majorana HNLs with existing constraints on $\mumix$ in Fig.~\ref{fig:limits} for $\lvert U_{e 4}\rvert^2 = \lvert U_{\tau 4}\rvert^2 = 0$.
Decays of stopped pions in the process
$\pi^+\to\mu^+\nu$ 
are sensitive to the range $\mhnl<m_\pi-m_\mu$. Such searches
have been performed at the Swiss National Institute (SIN) in the mass range $1<\mhnl<16$~MeV~\cite{Daum:1987bg} and by the PIENU Collaboration for 
$15.7<\mhnl<33.8$~MeV~\cite{PIENU:2019usb}. 
The muon spectrum measured in stopped $K^+\to\mu^+\nu$ decays $(K_{2\mu})$ has been used to set limits in the mass range
$70<\mhnl<300$~MeV with the E89 experiment at KEK~\cite{Hayano:1982wu}, in the range
$175<\mhnl<300$~MeV with the E949 experiment
at the Brookhaven National Laboratory (BNL)~\cite{E949_limit_2015},
and $200<\mhnl<384$~MeV by the NA62 experiment at CERN~\cite{NA62:2021bji}.
An update of the KEK-E89 result was reported in proceedings~\cite{Yamazaki:1984sj}, extending the sensitivity to $\mhnl>40$~MeV.
The PS191 experiment~\cite{PS191_1987} was specifically designed to search for massive decaying neutrinos in the CERN-PS proton beam. Its search for $N\to\nu\mu e$ decays constrains 
$\mumix$ in the range $\mhnl>m_{\mu}$.
Critical discussions of the PS191 results can be found in Refs.~\cite{Arguelles:2021dqn,Disputing_PS191_1,Ruchayskiy:2011aa,Kusenko:2004qc}.
The T2K Collaboration has published combined limits on $\mumix$, 
$\lvert U_{e 4}\rvert^2$, and $\lvert U_{\tau 4}\rvert^2$ for $\mhnl>150$~MeV in a Bayesian approach~\cite{Abe:2019kgx}.

In this letter, we use NuMI beam data recorded with the MicroBooNE detector to derive the most stringent constraints on $\mumix$ for the mass range $34<\mhnl<175$~MeV.
It is also the first search for HNLs in
$\vpi$ and $\ee$ final states using a LArTPC and the first ever result reported for HNL decays into $\vpi$.
When combined with the MicroBooNE limits on HNL decays into $\mpi$ pairs~\cite{MicroBooNE:2022ctm}, our results
now cover the full mass range $10<\mhnl<385$~MeV that is kinematically accessible from kaons produced by the NuMI beam.

This document was prepared by the MicroBooNE collaboration using the
resources of the Fermi National Accelerator Laboratory (Fermilab), a
U.S. Department of Energy, Office of Science, HEP User Facility.
Fermilab is managed by Fermi Research Alliance, LLC (FRA), acting
under Contract No. DE-AC02-07CH11359.  MicroBooNE is supported by the
following: 
the U.S. Department of Energy, Office of Science, Offices of High Energy Physics and Nuclear Physics; 
the U.S. National Science Foundation; 
the Swiss National Science Foundation; 
the Science and Technology Facilities Council (STFC), part of the United Kingdom Research and Innovation; 
the Royal Society (United Kingdom); 
the UK Research and Innovation (UKRI) Future Leaders Fellowship; 
and the NSF AI Institute for Artificial Intelligence and Fundamental Interactions. 
Additional support for 
the laser calibration system and cosmic ray tagger was provided by the 
Albert Einstein Center for Fundamental Physics, Bern, Switzerland. We 
also acknowledge the contributions of technical and scientific staff 
to the design, construction, and operation of the MicroBooNE detector 
as well as the contributions of past collaborators to the development 
of MicroBooNE analyses, without whom this work would not have been 
possible. For the purpose of open access, the authors have applied 
a Creative Commons Attribution (CC BY) public copyright license to 
any Author Accepted Manuscript version arising from this submission.

\bibliography{HNLee}

\end{document}

%% file: microboone-author-list-september2023-PRD.tex
\newcommand{\ANL}{Argonne National Laboratory (ANL), Lemont, IL, 60439, USA}
\newcommand{\Bern}{Universit{\"a}t Bern, Bern CH-3012, Switzerland}
\newcommand{\BNL}{Brookhaven National Laboratory (BNL), Upton, NY, 11973, USA}
\newcommand{\UCSB}{University of California, Santa Barbara, CA, 93106, USA}
\newcommand{\Cambridge}{University of Cambridge, Cambridge CB3 0HE, United Kingdom}
\newcommand{\CIEMAT}{Centro de Investigaciones Energ\'{e}ticas, Medioambientales y Tecnol\'{o}gicas (CIEMAT), Madrid E-28040, Spain}
\newcommand{\Chicago}{University of Chicago, Chicago, IL, 60637, USA}
\newcommand{\Cincinnati}{University of Cincinnati, Cincinnati, OH, 45221, USA}
\newcommand{\CSU}{Colorado State University, Fort Collins, CO, 80523, USA}
\newcommand{\Columbia}{Columbia University, New York, NY, 10027, USA}
\newcommand{\Edinburgh}{University of Edinburgh, Edinburgh EH9 3FD, United Kingdom}
\newcommand{\FNAL}{Fermi National Accelerator Laboratory (FNAL), Batavia, IL 60510, USA}
\newcommand{\Granada}{Universidad de Granada, Granada E-18071, Spain}
\newcommand{\Harvard}{Harvard University, Cambridge, MA 02138, USA}
\newcommand{\IIT}{Illinois Institute of Technology (IIT), Chicago, IL 60616, USA}
\newcommand{\Indiana}{Indiana University, Bloomington, IN 47405, USA}
\newcommand{\KSU}{Kansas State University (KSU), Manhattan, KS, 66506, USA}
\newcommand{\Lancaster}{Lancaster University, Lancaster LA1 4YW, United Kingdom}
\newcommand{\LANL}{Los Alamos National Laboratory (LANL), Los Alamos, NM, 87545, USA}
\newcommand{\Louisiana}{Louisiana State University, Baton Rouge, LA, 70803, USA}
\newcommand{\Manchester}{The University of Manchester, Manchester M13 9PL, United Kingdom}
\newcommand{\MIT}{Massachusetts Institute of Technology (MIT), Cambridge, MA, 02139, USA}
\newcommand{\Michigan}{University of Michigan, Ann Arbor, MI, 48109, USA}
\newcommand{\MSU}{Michigan State University, East Lansing, MI 48824, USA}
\newcommand{\Minnesota}{University of Minnesota, Minneapolis, MN, 55455, USA}
\newcommand{\Nankai}{Nankai University, Nankai District, Tianjin 300071, China}
\newcommand{\NMSU}{New Mexico State University (NMSU), Las Cruces, NM, 88003, USA}
\newcommand{\Oxford}{University of Oxford, Oxford OX1 3RH, United Kingdom}
\newcommand{\Pitt}{University of Pittsburgh, Pittsburgh, PA, 15260, USA}
\newcommand{\Rutgers}{Rutgers University, Piscataway, NJ, 08854, USA}
\newcommand{\SLAC}{SLAC National Accelerator Laboratory, Menlo Park, CA, 94025, USA}
\newcommand{\SDSMT}{South Dakota School of Mines and Technology (SDSMT), Rapid City, SD, 57701, USA}
\newcommand{\Maine}{University of Southern Maine, Portland, ME, 04104, USA}
\newcommand{\Syracuse}{Syracuse University, Syracuse, NY, 13244, USA}
\newcommand{\TelAviv}{Tel Aviv University, Tel Aviv, Israel, 69978}
\newcommand{\Tennessee}{University of Tennessee, Knoxville, TN, 37996, USA}
\newcommand{\UTA}{University of Texas, Arlington, TX, 76019, USA}
\newcommand{\Tufts}{Tufts University, Medford, MA, 02155, USA}
\newcommand{\UCL}{University College London, London WC1E 6BT, United Kingdom}
\newcommand{\VTech}{Center for Neutrino Physics, Virginia Tech, Blacksburg, VA, 24061, USA}
\newcommand{\Warwick}{University of Warwick, Coventry CV4 7AL, United Kingdom}
\newcommand{\Yale}{Wright Laboratory, Department of Physics, Yale University, New Haven, CT, 06520, USA}

\affiliation{\ANL}
\affiliation{\Bern}
\affiliation{\BNL}
\affiliation{\UCSB}
\affiliation{\Cambridge}
\affiliation{\CIEMAT}
\affiliation{\Chicago}
\affiliation{\Cincinnati}
\affiliation{\CSU}
\affiliation{\Columbia}
\affiliation{\Edinburgh}
\affiliation{\FNAL}
\affiliation{\Granada}
\affiliation{\Harvard}
\affiliation{\IIT}
\affiliation{\Indiana}
\affiliation{\KSU}
\affiliation{\Lancaster}
\affiliation{\LANL}
\affiliation{\Louisiana}
\affiliation{\Manchester}
\affiliation{\MIT}
\affiliation{\Michigan}
\affiliation{\MSU}
\affiliation{\Minnesota}
\affiliation{\Nankai}
\affiliation{\NMSU}
\affiliation{\Oxford}
\affiliation{\Pitt}
\affiliation{\Rutgers}
\affiliation{\SLAC}
\affiliation{\SDSMT}
\affiliation{\Maine}
\affiliation{\Syracuse}
\affiliation{\TelAviv}
\affiliation{\Tennessee}
\affiliation{\UTA}
\affiliation{\Tufts}
\affiliation{\UCL}
\affiliation{\VTech}
\affiliation{\Warwick}
\affiliation{\Yale}

\author{P.~Abratenko} \affiliation{\Tufts}
\author{O.~Alterkait} \affiliation{\Tufts}
\author{D.~Andrade~Aldana} \affiliation{\IIT}
\author{L.~Arellano} \affiliation{\Manchester}
\author{J.~Asaadi} \affiliation{\UTA}
\author{A.~Ashkenazi}\affiliation{\TelAviv}
\author{S.~Balasubramanian}\affiliation{\FNAL}
\author{B.~Baller} \affiliation{\FNAL}
\author{G.~Barr} \affiliation{\Oxford}
\author{D.~Barrow} \affiliation{\Oxford}
% ...

\collaboration{The MicroBooNE Collaboration}
\thanks{microboone\_info@fnal.gov}\noaffiliation